# Second-order Overtone and Combinational Raman Modes of Graphene Layers in the Range of 1690 cm$^{-1}$ to 2150 cm$^{-1}$

Chunxiao Cong[†], Ting Yu*[†§], Riichiro Saito[‡]

[†]Division of Physics and Applied Physics, School of Physical and Mathematical Sciences, Nanyang Technological University, 637371, Singapore

[§]Department of Physics, Faculty of Science, National University of Singapore, 117542, Singapore

[‡]Department of Physics, Tohoku University, Sendai, Miyagi 9808578 Japan

*To whom correspondence should be addressed Email: yuting@ntu.edu.sg

**ABSTRACT**

Though graphene has been intensively studied by Raman spectroscopy, in this letter, we report a study of second-order overtone and combinational Raman modes in an *unexplored* range of 1690-2150 cm$^{-1}$ in nonsuspended commensurate (AB-stacked), incommensurate (folded) and suspended graphene layers. Based on the double resonance theory, four dominant modes in this range have been assigned as 2oTO (M band), iTA+LO, iTO+LA and LO+LA. Differing to AB-stacked bilayer graphene or few layer graphene, the M band disappears in single layer graphene. Systematic analysis reveals that interlayer interaction is essential for the presence (or absence) of M band whereas the substrate has no effect on this. Dispersive behaviors of these "new" Raman modes in graphene have been probed by the excitation energy dependent Raman spectroscopy. It is found that the appearance of the M band strictly relies on the AB stacking, which could be a fingerprint of AB-stacked bilayer graphene. This work expands the unique and powerful abilities of Raman spectroscopy on study of graphene and provides another effective way to probe phonon dispersion, electron-phonon coupling, and to exploit electronic band structure of graphene layers.

**KEYWORDS**: Graphene, Raman, AB stacking, Electron-phonon coupling

Carbon materials occurring in many different forms, such as highly ordered pyrolytic graphite (HOPG), diamond, carbon fibers, carbon nanotubes, buckminsterfullerene and so on, are very important for science and technology.[1, 2] The recently discovered single layer graphene (SLG), which has unusual electronic properties, shows remarkable signs of applicability for fundamental studies and applications in nanoelectronics and nanophotonics.[3-5] Moreover, it was recently shown, both experimentally [6-8] and theoretically [9, 10] that the electronic gap between the valence and conduction bands of AB-stacked bilayer graphene (BLG) can be controlled by an applied electric field. This makes BLG the only known semiconductor with a tunable energy gap and may open the way for developing photo detectors and lasers tunable by the electric field effect.[6] Thus growing large area of BLG is imperative for the real application and many researchers have devoted great efforts on this. Therefore, a fast accurate and nondestructive method, which can be used to identify the AB-stacked BLG from others, is in urgent need.

In this work, we exploited the second-order overtone and combinational Raman modes in the range of 1690-2150 cm$^{-1}$ of graphene layers and identified all the dominant modes in this range based on the double resonance theory.[11, 12] It is found that a band at about 1750 cm$^{-1}$, assigned as M band only appears in the BLG or few layer graphene (FLG) and is absent in SLG and incommensurate bilayer graphene (IBLG). This provides a fast and reliable way to identify the BLG from SLG and IBLG, which must be very useful for the future study on the BLG and development of nanodevices based on BLG.

As exploited by the previous experimental and theoretical studies, the fantastic properties of graphene layers are due to their unique electronic band structures. Understanding the behavior of electrons and phonons in graphene layers is also very important for their practical applications of electronic devices. Raman spectra provide precise information on the crystal structures, electronic band structures, the phonon energy dispersion, and the electron-phonon interaction in sp2 carbon systems.[13] It has been shown experimentally that by monitoring the position, width, integrated intensity, shape and so on of Raman features of graphene layers, the number of layers, [14, 15] linear dispersion of electronic energy, [16] crystal orientation, [17-19] doping, [20, 21] defects [22] and strain [23, 24] can be probed. The most prominent peaks in the Raman spectrum of graphene layers are the so called G mode and G' (or 2D) mode, which are associated with the doubly degenerate (iTO and iLO) phonon mode at Γ point and with inter-valley double resonance process involving two iTO phonons near *K* point, respectively. If laser beam is focused on edges of graphene or some defects exist, another two peaks will appear, named D mode and D' mode, which are inter-valley double resonance process involving defects and iTO phonons and intra-valley double resonance process involving defects and iLO phonons. To our knowledge, Raman features in the range of 1690-2150 cm$^{-1}$ in graphene layers were never reported, though they have been studied in single wall carbon nanotubes (SWNTs) [25, 26] and highly ordered pyrolytic graphite (HOPG),[25] graphite whiskers,[27] double wall carbon nanotubes (DWCNTs)[28] and multiwall carbon nanotubes (MWNTs).[29] Our works presented in the letter disclose the nature of these modes. The energy dependence of the combinational modes in this range is given through a detailed excitation energy dependent Raman spectroscopy study.

In this work, the nonsuspendend commensurate (AB-stacked), incommensurate (folded) and suspended graphene layers were studied. The graphene layers were prepared by mechanical cleavage of HOPG and transferred onto a 300 nm SiO$_2$/Si substrate. The majority of the transferred graphene layers are AB-stacked whereas some parts of the SLG may fold over during the transfer and form incommensurate BLG.[30] The detailed process of fabrication suspended samples was described in Ref. 31. In general, a 300 nm SiO$_2$/Si substrate with pre-patterned holes was used. After transfer, some parts of graphene layers covering the holes are suspended. An optical microscope was used to locate the thin layers, and the number of layers was further identified by white light contrast spectra, width of G' mode and absolute Raman intensity of G mode. The white light contrast spectra were acquired using a WITec CRM200 Raman system with a 150 lines/mm grating. The Raman spectra were obtained using a Renishaw system with a 2400 lines/mm grating cooperated with a 532 nm laser ($E_{laser}$ = 2.33 eV). The excitation energy dependent Raman spectra were recorded using a WITec CRM200 Raman system with a 600 lines/mm grating. The laser power was kept below 0.1 mW on the sample surface to avoid laser induced heating.

Properties of graphene layers could be significantly different when their thicknesses or number of layers are varying. Several techniques have been adapted for identifying number of layers of graphene films, such as quantum Hall effect measurement, [4,5] atomic force microscopy (AFM), [5] transmission electron microscopy (TEM), [32] Raman spectroscopy [14] and contrast spectroscopy [15] etc. Among them, the optical approaches like Raman and contrast spectroscopy are most favorable. Figure 1(a) shows the optical images of selected graphene layers. The white light contrast spectra of the selected graphene layers are shown in Fig. 1(b). By using the method reported in Ref. 15, the number of layers of these graphene films is determined from one to seven, as labeled in the optical images. The extract contrast value as a function of number of layers shown in Fig. 1(c) is also in agreement with the results in Ref. 15. To double confirm the number of layers, Raman spectra of these graphene films were recorded and fitted. Fig. 1(d) presents the G band intensity as a function of number of layers. The linear dependence could be another inspection of layer determination.[33] The width of G' mode was also measured to identify the number of layers of graphene films.[14]

Graphene has been intensively studied by Raman spectroscopy. When researchers enjoy the beauty of the strong resonant modes like D, G, and G', very little attention has been afforded to the range of 1690 cm$^{-1}$ to 2150 cm$^{-1}$, where actually several modes have been observed in other graphitic materials like carbon nanotubes, graphite and graphite whisker. Equally importantly and interestingly, these modes are also resulted from the double resonance process and carry information of electron-phonon coupling. Figure 2 shows the Raman spectra of the graphene layers in this range together with the spectrum of HOPG. As comparison, the G' mode is also shown in Fig. 2(c). Similarly, but much more obviously compared to the G' mode, the Raman modes between 1690 cm$^{-1}$ and 2150 cm$^{-1}$ display an evolution of the spectral components such as line-shape, peak position and peak width with the increment of the number of layers.

To reveal the nature of these modes within this unexplored range, we divide them into two groups. One is in the range of 1690 cm$^{-1}$ to 1800 cm$^{-1}$ (Fig. 2(a)) and the other is in the range of 1800 cm$^{-1}$ to 2150 cm$^{-1}$ (Fig. 2(b)), which corresponds to the M band range and combinational mode range as discussed previously for SWNTs.[25] In the M band range, the most remarkable observation is the absence of any peak in the spectrum of SLG, whereas the BLG and FLG show two obvious peaks with multiple components. This absolutely differs from the G' mode, which shows an asymmetric wider peak due to the extra components led by the existence of extra electronic branches in BLG, while only one sharp and strong peak presented in the spectrum of SLG. This distinguished behavior of the modes in the M band range and G' mode could be explained by their own natures. G' mode or the overtone of the iTO mode is Raman active for all graphene layers. In contrast, the M band is the overtone of the oTO phonon, an infrared-active out-of-plane mode. For SWNTs due to the confinement (zone folding) effects and for HOPG due to the interlayer coupling, the Raman mode selection rule could be relaxed and this oTO mode becomes Raman active. In the case of SWNTs, depending on the energy of the van Hove singularity relative to the excitation energy, the M band could selectively appear in some carbon nanotubes of certain chirality ($n, m$) and diameter.[34] Furthermore, for SWNTs, various line-shapes and a steplike phonon dispersion of such intermediate frequency mode could be also resulted from this dependence.[35] On the other hand, for HOPG, FLG and BLG, there is no such dependence. The resonance condition could be always satisfied. The non-zone-center phonons connected with the Γ point reflected by this oTO mode undergo the electron double resonant intravelly scattering in two means, one with a near-zero momentum, appearing as $M^+$ band, and the other with a momentum of the double of that of the scattered excited electrons, named $M^-$ band (indicated in Fig. 2(a)).[25] Considering the above truth, we attribute the mode of BLG and FLG in the range of 1690-1800 $cm^{-1}$ to the M band caused by an intravalley double resonance scattering process. Similar to the G' mode, the evolution of this M band with the increase of the number of layers might be mainly due to the evolution of electronic band structure of graphene films. The existence of this M band could be a fingerprint of the AB stacking BLG. More discussion will be given below.

The evolution of the modes could be also observed in the spectra in the range of 1800 $cm^{-1}$ to 2150 $cm^{-1}$ as shown in Fig. 2(b). Several arguments about the assignments of the Raman modes of carbon materials in this range were reported in previous studies. For example, in HOPG (SWNTs), the peak at 1950 $cm^{-1}$ (1987 $cm^{-1}$) was tentatively assigned as a combination of the in-plane transverse optic (iTO) and longitudinal acoustic (LA) modes (iTO+LA).[25] Another group claimed that this asymmetric peak consists of two modes, named iTO+LA and LO+LA.[26] For graphite whiskers, the two peaks in this range were identified as iTA+LO and LO+LA.[27] However, no conclusive assignments are available, especially for graphene. With careful curve fitting and systematic analysis, we believe there are three dominant peaks in this combinational mode range.

Figure 3(a) shows the Raman spectra of the SLG in the combinational mode range with various excitation energies. Very remarkably, all the peaks display a significant blue-shift with the increase of the excitation energy. To exploit the details, the experimental data were fitted by three Lorentzian line shape peaks and the fitting curves match the experimental data perfectly. The frequencies of these modes were plotted as a function of the excitation energies. A positive and linear excitation energy dependence of these phonons are clearly presented in Fig. 3(b), which indicates the origin of these peaks are double resonance process, the same as the well-studied D and G' modes. To fulfill the double resonance Raman process, the momentum of the phonon must be near the double of that of the intravalley or intervalley scattered electrons. In SLG, at relative low energy level (<3 eV), the electronic spectrum is linear near the *K* point. Therefore, the electron energy shows a linear relationship with the phonon momentum, which promises Raman spectroscopy to be a unique and powerful tool for probing both electronic structure and phonon dispersion of graphene. Benefiting from this linear dependence, we depict the phonon dispersion frequency versus phonon wave vector in Fig. 3(b). The energy dependence (the slope) of these three peaks is estimated to be 140 cm$^{-1}$/eV for the lowest frequency mode, 198 cm$^{-1}$/eV for the medium frequency one and 221 cm$^{-1}$/eV for the highest frequency mode. For the 532 nm laser or excitation energy of 2.33 eV, the lowest frequency mode locates at 1860 cm$^{-1}$, which perfectly matches the combination of iTA and iTO. (Our readings from the dispersion curve in Ref. 25 are that $\omega_{iTA}$ = 300 cm$^{-1}$ and $\omega_{iTO}$ = 1560 cm$^{-1}$ for 2.33eV). However, the frequency of a combinational mode is usually a bit less than the sum frequency of the mother-components.[27] Moreover, the energy dependence or the slope of the dispersion curve of the iTA mode is 129 cm$^{-1}$/eV [27] and the iTO mode shows a negative dispersion in the range studied in this work, which cannot lead a mode with an energy dependence of 140 cm$^{-1}$/eV by combining these two modes (iTA+iTO). In contrast to the iTO mode, the LO mode positively disperses within this range and locates at 1600 cm$^{-1}$ for 2.33eV laser. Therefore, the iTA+LO could be responsible for this low frequency peak. This assignment could also be supported by the study on graphite whiskers, where an energy dependence of 139 cm$^{-1}$/eV was observed for an iTA+LO mode.[27] This perfect 'agreement' may hint that the SLG may share some structural commons with the graphite whiskers. Following the same strategy and considering the truth that the energy dependence of the LA mode is 216 cm$^{-1}$/eV, [27] the medium frequency mode with an energy dependence of 198 cm$^{-1}$/eV is attributed to a combination of iTO (with a negative energy dependence) and LA, while the high frequency mode with an energy dependence of 221 cm$^{-1}$/eV is attributed to a combination of LO (with a positive energy dependence) and LA. This agrees with the work on SWNTs [26] and could explain why the peak shows an asymmetric line shape in Ref. 27, which actually consists of two combinational modes: iTO+LA and LO+LA.

As a single atomic layer, graphene is extremely sensitive to the substrates, which has been clearly reflected in the Raman spectra.[24] To investigate the substrate effects on the Raman spectra of graphene in the range of 1690 cm$^{-1}$ and 2150 cm$^{-1}$, the non-suspended and suspended SLG and BLG were prepared and the corresponding spectra are shown in Fig. 4(a). From the spectra, it can be clearly seen that the substrate has no significant effects on the M band and the combinational modes of SLG and BLG. The further analysis of the interlayer coupling was performed on the Raman spectrum of the IBLG. Similar to the G' mode, the integrated intensity of the combinational modes is dramatically increased in such IBLG, and a blue-shift presents, which is due to the reduction of charging impurities.[36] The most significant and meaningful observation is the twisting of the two layers or the breaking of the AB (Bernal) stacking totally suppresses the M band shown in Raman spectrum of the IBLG. This might be due to the breaking of the stacking order of the A, B sub-lattice in two layers and consequently the oTO mode becomes non-Raman active mode. The IBLG is more like two 'isolated' SLG. Detailed theoretical study is needed to exploit the interlayer coupling effects on this M band. The missing of M band could be also observed in the Raman spectrum of the graphite whisker, which is similar as disordered graphite or thick incommensurate graphene layers,[27] once again, indicating the commons among graphite whisker, IBLG and SLG. The absence of the M band in the IBLG makes the M band a unique feature for identifying the AB-stacked BLG, which possesses plenty of practical potentials.

In summary, we have studied the second order overtone and combinational Raman modes of graphene layers in an unexplored range of 1690-2150 cm$^{-1}$. All the dominant modes in this range have been assigned based on the double resonance Raman process. The energy dependence of the combinational modes in the visible range has been derived from the excitation energy dependent Raman spectroscopy. The M band, which is strictly relayed on the AB stacking, provides a fast and accurate way to identify the BLG from SLG and IBLG. All the modes in this range show the layer dependent evolution, which should be relative to the changes of the electronic structures of the graphene layers. The results herein certainly demonstrate the unique and new potentials of Raman scattering for probing electronic band structures, the phonon energy dispersion and electron-phonon interaction in graphene layers.

*Acknowledgements*

This work is supported by the Singapore National Research Foundation under NRF RF Award No. NRF-RF2010-07 and MOE Tier 2 MOE2009-T2-1-037. Cong thanks Professor Toshiaki Enoki for his enlightening discussion. RS acknowledges MEXT Grant (No. 20241023).

**Reference**

[1] Dresselhaus, M. S.; Dresselhaus, G.; Sugihara, K.; Spain, I. L.; Goldberg, H. A.; Graphite Fibers and Filaments, **1988,** Vol. 5 of Springer Series in Materials Science (Springer-Verlag, Berlin).

[2] Ollle Ingan; Ingemar Lundst Carbon Nanotube Muscles. Science **1999**, 284, 1281-1282.

[3] Novoselov, K. S.; Geim, A. K.; Morozov, S. V.; Jiang, D.; Zhang, Y.; Dubonos, S. V.; Grigorieva, I. V.; Firsov, A. A. Electric Field Effect in Atomically Thin Carbon Films. Science, **2004**, 306, 666-669.

[4] Novoselov, K. S.; Geim, A. K.; Morozov, S. V.; Jiang, D.; Katsnelson, M. I.; Grigorieva, I. V.; Dubonos, S. V.; Firsov, A. A. Two-dimensional gas of massless Dirac fermions in graphene. Nature, **2005**, 438, 197-200.

[5] Zhang, Y.; Tan, Y. W.; Stormer, H. L.; Kim, P. 1.Experimental observation of the quantum Hall effect and Berry's phase in graphene. Nature, **2005**, 438, 201-204.

[6] Castro, E. V.; Novoselov, K. S.; Morozov, S. V.; Peres, N. M. R.; Lopes dos Santos, J. M. B.; Nilsson, J.; Guinea, F.; Geim, A. K.; Castro Neto, A. H. Phys. Rev. Lett. **2007**, 99, 216802.

[7] Oostinga, J. B.; Heersche, H. B.; Liu, X.; Morpurgo, A. F.; Vandersypen, L. M. K. Gate-induced insulating state in bilayer graphene devices. Nat. Mater. **2007**, 7, 151-157.

[8] Zhang, Y.; Tang, T.-T.; Girit, C.; Hao, Z.; Martin, M. C.; Zettl, A.; Crommie, M. F.; Shen, Y. R.; Wang, F. 1.Direct observation of a widely tunable bandgap in bilayer graphene. Nature, **2009**, 459, 820-823.

[9] McCann, E. Asymmetry gap in the electronic band structure of bilayer graphene. Phys. Rev. B 74, **2006**, 161403(R)-161406(R).

[10] Min, H.; Sahu, B.; Banerjee, S. K.; MacDonald, A. H. Ab initio theory of gate induced gaps in graphene bilayers. Phys. Rev. B **2007**, 75, 155115-155121.

[11] Thomsen, C.; Reich, S. Double Resonant Raman Scattering in Graphite. Phys. Rev. Lett. **2000**, 85, 5214-5217.

[12] Saito, R.; Jorio, A.; Souza Filho, A. G.; Dresselhaus, G.; Dresselhaus, M. S.; Pimenta, M. A. Probing Phonon Dispersion Relations of Graphite by Double Resonance Raman Scattering. Phys. Rev. Lett. **2002**, 88, 027401-1-027401-4.

[13] Dresselhaus, M. S.; Jorio, A.; Saito, R. Characterizing Graphene, Graphite, and Carbon Nanotubes by Raman Spectroscopy. Annu. Rev. Condens. Matter Phys. **2010**, 1, 89-108.

[14] Ferrari, A. C.; Meyer, J. C.; Scardaci, V.; Casiraghi, C.; Lazzeri, M.; Mauri, F.; Piscanec, S.; Jiang, D.; Novoselov, K. S.; Roth, S.; *et al.* Raman Spectrum of Graphene and Graphene Layers Phys. Rev. Lett. **2006**, 97, 187401-1-187401-4.

[15] Ni, Z. H.; Wang, H. M.; Kasim, J.; Fan, H. M.; Yu, T.; Wu, Y. H.; Feng, Y. P.; Shen, Z. X. Graphene Thickness Determination Using Reflection and Contrast Spectroscopy. Nano Lett. **2007**, 7, 2758-2763.

[16] Mafra, D. L.; Samsonidze, G.; Malard, L. M.; Elias, D. C.; Brant, J. C.; Plentz, F.; Alves, E. S.; Pimenta, M. A. Determination of LA and TO phonon dispersion relations of graphene near the Dirac point by double resonance Raman scattering. Phys. Rev. B **2007**, 76, 233407-1-233407-4.

[17] Huang, M.; Yan, H.; Chen, C.; Song, D.; Heinz, T. F.; Hone, J. Phonon Softening and Crystallographic Orientation of Strained Graphene Studied by Raman Spectroscopy. PNAS **2009**, 106, 7304-7308.

[18] Mohiuddin, T. M. G.; Lombardo, A.; Nair, R. R.; Bonetti, A.; Savini, G.; Jalil, R.; Bonini, N.; Basko, D. M.; Galiotis, C.; Marzari, N.; et al. Uniaxial Strain in Graphene by Raman Spectroscopy: G Peak Splitting, Grüneisen Parameters, and Sample Orientation. Phys. Rev. B **2007**, 79, 205433-1-205433-8.

[19] You, Y. M.; Ni, Z. H.; Yu, T.; Shen, Z. X. Edge Chirality Determination of Graphene by Raman Spectroscopy. Appl. Phys. Lett. **2008**, 93, 163112-1-163112-3.

[20] Das, A.; Pisana, S.; Chakraborty, B.; Piscanec, S.; Saha, S. K.; Waghmare, U. V.; Novoselov, K. S.; Krishnamurthy, H. R.; Geim, A. K.; Ferrari, A. C.; et al. Monitoring Dopants by Raman Scattering in an Electrochemically Top-Gated Graphene Transistor. Nat. Nanotechnol. **2008**, 3, 210-215.

[21] Luo, Z.; Yu, T,; Kim, K.; Ni, Z.; You, Y.; Lim, S.; Shen, Z.; Wang, S.; Lin, J. Thickness-Dependent Reversible Hydrogenation of Graphene Layers. ACS Nano **2009**, 3, 1781-1788.

[22] Cancado, L. G.; Pimenta, M. A.; Saito, R.; Jorio, A.; Ladeira, L. O.; Grueneis, A.; Souza-Filho, A. G.; Dresselhaus, G.; Dresselhaus, M. S. Stokes and Anti-Stokes Double Resonance Raman Scattering in Two-Dimensional Graphite. Phys. Rev. B **2002**, 66, 035415-1-035415-5.

[23] Yu, T.; Ni, Z.; Du, C.; You, Y.; Wang, Y.; Shen, Z. Raman Mapping Investigation of Graphene on Transparent Flexible Substrate: The Strain Effect. J. Phys. Chem. C **2008**, 112, 12602-12605.

[24] Ni, Z. H.; Yu, T.; Lu, Y. H.; Wang, Y. Y.; Feng, Y. P.; Shen, Z. X. Uniaxial Strain on Graphene: Raman Spectroscopy Study and Band-Gap Opening. ACS Nano **2008**, 2, 2301-2305.

[25] Brar, V. W.; Samsonidze, Ge. G.; Dresselhaus, M. S.; Dresselhaus, G.; Saito, R.; Swan, A. K.; Ünlü, M. S.; Goldberg, B. B.; Souza Filho, A. G.; Jorio, A. Second-order harmonic and combination modes in graphite, single-wall carbon nanotube bundles, and isolated single-wall carbon nanotubes. Phys. Rev. B **2002**, 66, 155418-155427.

[26] Fantini, C.; Pimenta, M. A.; Strano, M. S. Two-Phonon Combination Raman Modes in Covalently Functionalized Single-Wall Carbon Nanotubes. J. Phys. Chem. C **2008**, 112, 13150-13155.

[27] Tan, P. H.; Dimovski, S.; Gogotsi, Y. Raman scattering of non-planar graphite: arched edges, polyhedral crystals, whiskers and cones. Phil. Trans. R. Soc. Lond. A **2004**, 362, 2289-2310.

[28] Ellis, A. V. Second-order overtone and combination modes in the LOLA region of acid treated double-walled carbon nanotubes. J. Chem. Phys. **2006**, 125, 121103-1-121103-5.

[29] Tan, P. H.; An, L.; Liu, L. Q.; Guo, Z. X.; Czerw, R.; Carrol, D. L.; Ajayan, P. M.; Zhang, N.; Guo, H. L. Probing the phonon dispersion relations of graphite from the double-resonance process of Stokes and anti-Stokes Raman scatterings in multiwalled carbon nanotubes. Phys. Rev. B **2002**, 66, 245410-1-245410-8.

[30] Ni, Z. H.; Liu, L.; Wang, Y. Y.; Zheng, Z.; Li, L-J.; Yu, T.; Shen, Z. X.; G-band Raman Double Resonance in Twisted Bilayer Graphene: Evidence of Band Splitting and Folding. Phys. Rev. B **2009**, 80, 125404-1-125404-5.

[31] Ni, Z. H.; Yu, T.; Luo, Z. Q.; Wang, Y. Y.; Liu, L.; Wong, C. P.; Miao, J.; Huang, W.; Shen, Z. X. Probing Charged Impurities in Suspended Graphene Using Raman Spectroscopy. ACS Nano **2009**, 3, 569-574.

[32] Meyer, Jannik C.; Geim, A. K.; Katsnelson, M. I.; Novoselov, K. S.; Booth, T. J.; Roth, S. The structure of suspended graphene sheets. Nature, **2007**, 446, 60-63.

[33] Graf, D.; Molitor, F.; Ensslin, K.; Stampfer, C.; Jungen, A.; Hierold, C.; Wirtz, L. Spatially Resolved Raman Spectroscopy of Single- and Few-Layer Graphene. Nano Lett. **2007**, 7, 238-242.

[34] Samsonidze, Ge. G.; Saito, R.; Jorio, a.; Souza Filho, A. G.; Grüneis, A.; Pimenta, M. A.; Dresselhaus, G.; Dresslhaus, M. S. Phonon Trigonal Warping Effect in Graphite and Carbon Nanotubes. Phys. Rev. Lett. **2003**, 90, 027403-1-027403-4.

[35] Fantini, C.; Jorio, A.; Souza, M.; Saito, R.; Samsonidze, Ge. G.; Dresselhaus, M. S.; Pimenta, M. A. Steplike Dispersion of the Intermediate-frequency Raman Modes in Semiconducting and Metallic Carbon Nanotubes. Phys. Rev. B **2005**, 72, 085446-1-085446-5.

[36] Ni, Z. H.; Wang, Y. Y; Yu, T.; You, Y.M.; Shen, Z. X. Reduction of Fermi velocity in folded graphene observed by resonance Raman spectroscopy. Phys. Rev. B **2008**, 77, 205403-1-205403-5.

**Figure captions**

**Figure 1**. (a) Optical images of the selected graphene films with different numbers of layers. (b) The contrast spectra of the selected graphene films as indicated in (a). (c) The contrast value extracted from (b) as a function of number of layers. (d) G band absolute integrated intensity of graphene films measured under the same condition.

**Figure 2.** Raman spectra of a 300 nm SiO$_2$/Si substrate, graphene sheets with different numbers of layers and HOPG supported on the 300 nm SiO$_2$/Si substrate in the range of (a) 1690-1800 cm$^{-1}$, (b) 1800-2150 cm$^{-1}$ and (c) 2550-2850 cm$^{-1}$.

**Figure 3**. Raman spectra taken at several $E_{laser}$ values for SLG in the combinational mode range.

**Figure 4.** (a) Raman spectra of the suspended SLG and BLG, and the non-suspended SLG, BLG, and the incommensurate BLG. (b), (c) and (d) are optical images of the incommensurate BLG, the suspended BLG and the suspended SLG, respectively.

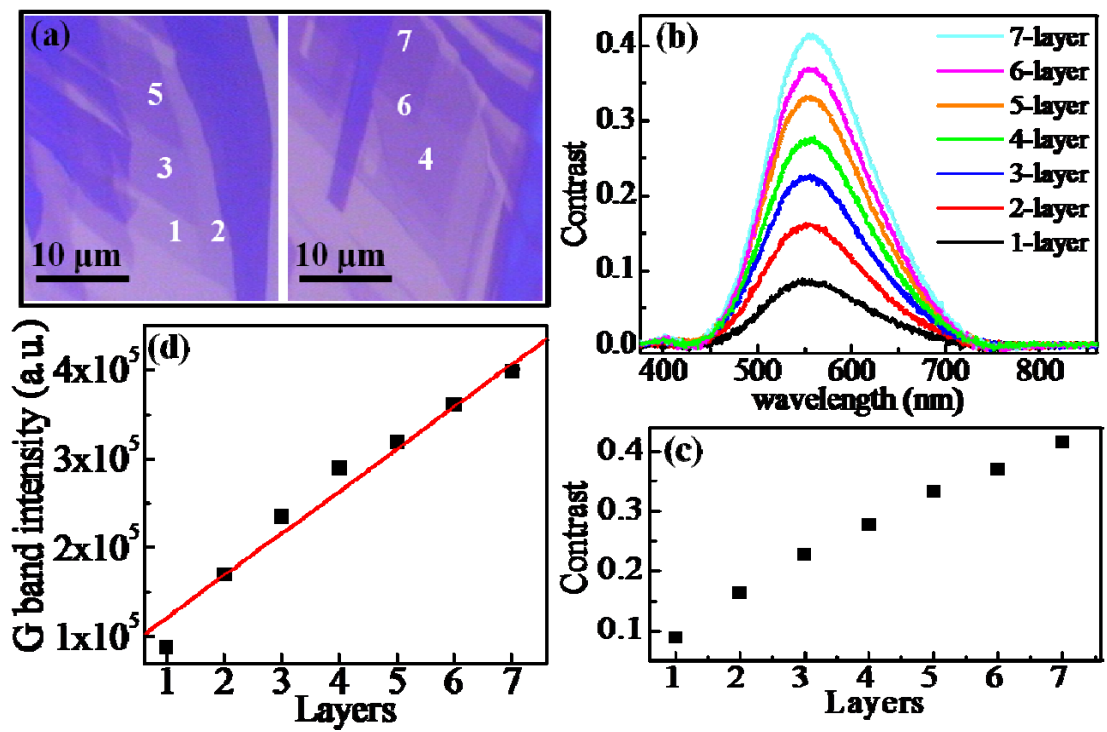

**Figure 1**

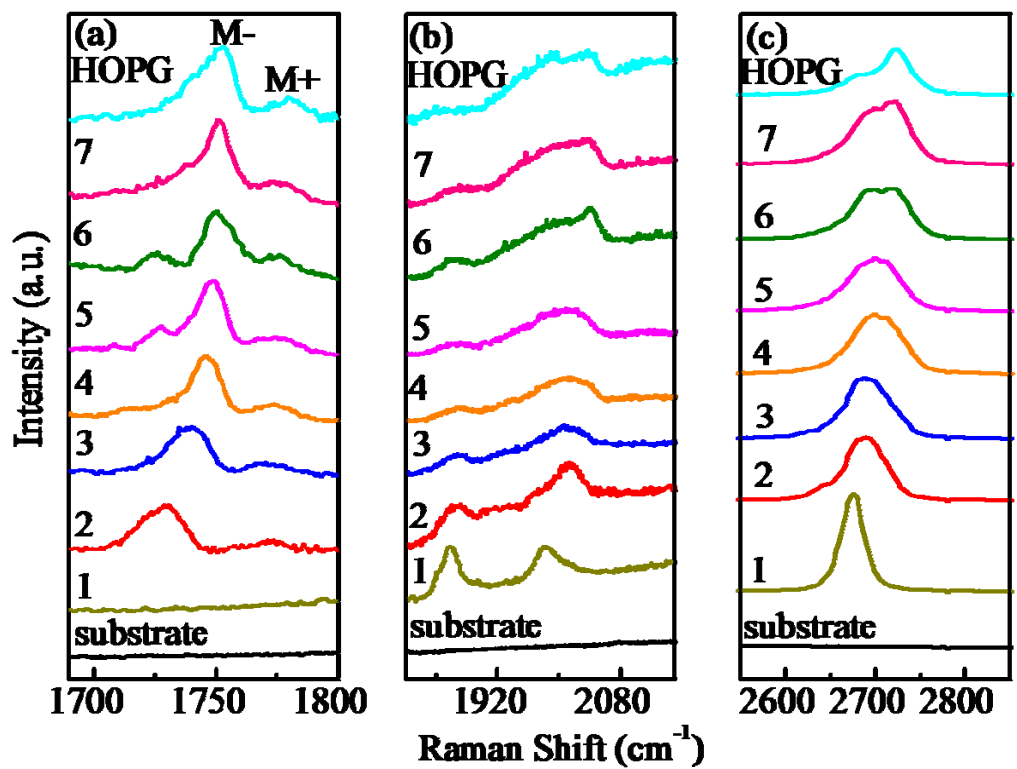

**Figure 2**

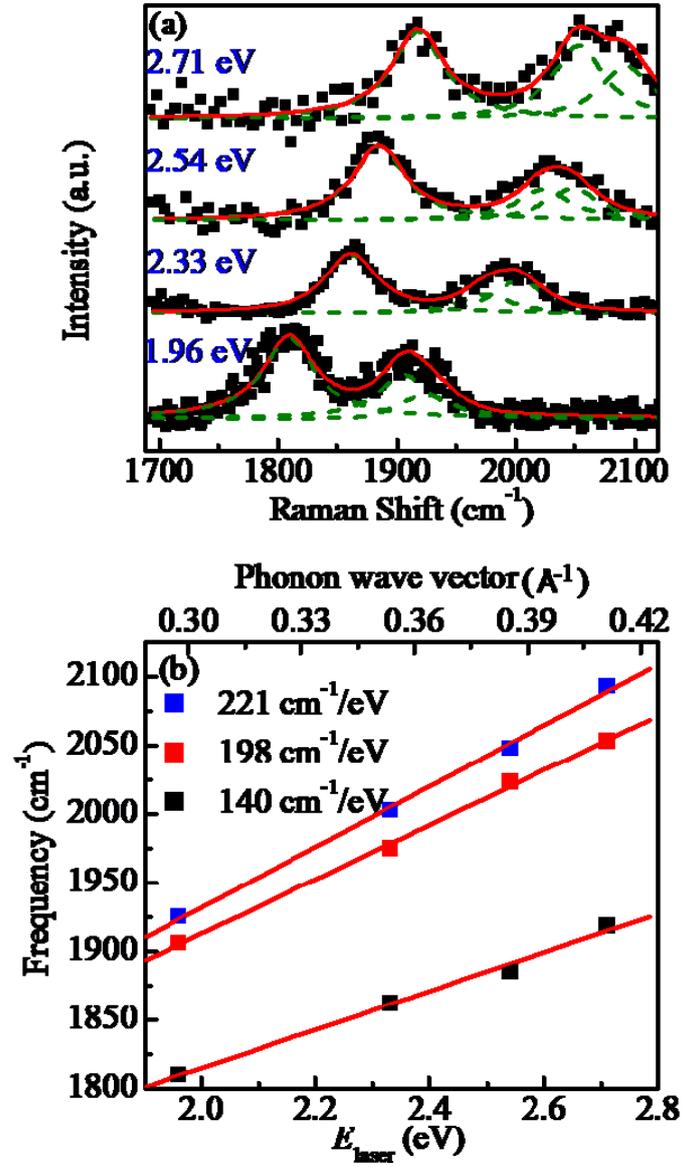

Figure 3

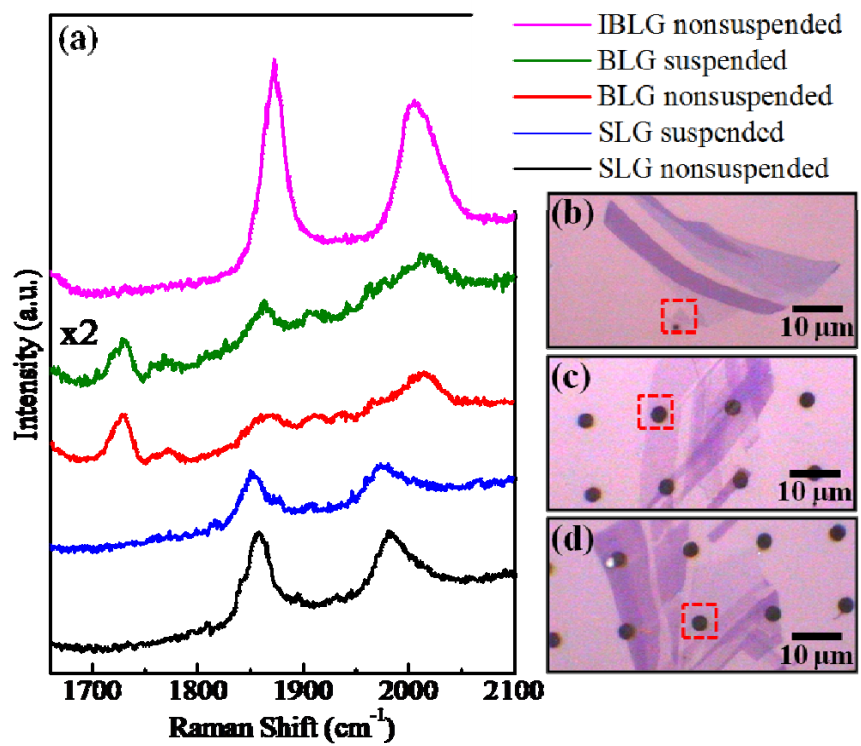

**Figure 4**